\documentclass[useAMS,usegraphicx,usenatbib]{mn2e}

\newcommand{\Msun}{\ensuremath{\,{\rm M}_\odot}}           
\newcommand{\kms}{\,km\,s$^{-1}$}                          
\newcommand{\ion}[2]{{#1}\,{\sc {\small{#2}}}}             
\newcommand{\Porb}{\ensuremath{P_{\rm orb}}}               
\newcommand{\mc}[1]{\multicolumn{2}{c}{#1}}
\newcommand{\cd}{\ensuremath{\,{\rm cycle\,\,d}^{-1}}}     
\newcommand{\as}{\ensuremath{^{\prime\prime}}}             
\newcommand{\reff}[1]{{\bf #1}}                                  

\title[Orbital periods of SDSS cataclysmic variables. II.]
      {Orbital periods of cataclysmic variables identified by the SDSS. II.
       Measurements for six objects, including two eclipsing systems}

\author[Southworth et al.]
       {John Southworth$^1$%
        \thanks{E-mail: j.k.taylor@warwick.ac.uk (JS),
        \newline T.R.Marsh@warwick.ac.uk (TRM),
        \newline Boris.Gaensicke@warwick.ac.uk (BTG)},
        T.\ R.\ Marsh$^1$,
        B.\ T.\ G\"ansicke$^1$,
        A.\ Aungwerojwit$^{1,2}$, \newauthor
        P.\ Hakala$^3$,
        D.\ de Martino$^4$,
        H.\ Lehto$^{3,5}$ \\
        $^1$ Department of Physics, University of Warwick, Coventry, CV4 7AL, UK                            \\
        $^2$ Department of Physics, Faculty of Science, Naresuan University, Phitsanulok, 65000, Thailand   \\
        $^3$ Tuorla Observatory, University of Turku, FIN-21500 Piikki\"o, Finland                          \\
        $^4$ INAF - Osservatorio di Capodimonte, Via Moiariello 16, 80131 Napoli, Italy                     \\
        $^5$ Department of Physics, FIN-20014 University of Turku, Finland                                  }

\begin{document} \maketitle 

\begin{abstract}
Continuing our work from Paper\,I (Southworth et al., 2006) we present medium-resolution spectroscopy and broad-band photometry of seven cataclysmic variables (CVs) discovered by the Sloan Digital Sky Survey. For six of these objects we derive accurate orbital periods, all which are measured for the first time. For SDSS\,J013132.39$-$090122.2, which contains a non-radially pulsating white dwarf, we find an orbital period of $81.54 \pm 0.13$\,min and a low radial velocity variation amplitude indicative of an extreme mass ratio. For SDSS\,J205914.87$-$061220.4, we find a period of $107.52 \pm 0.14$\,min. This object is a dwarf nova and was fading from its first recorded outburst throughout our observations. INT photometry of SDSS\,J155531.99$-$001055.0 shows that this system undergoes total eclipses which are 1.5\,mag deep and occur on a period of $113.54 \pm 0.03$\,min. A NOT light curve of SDSS\,J075443.01$+$500729.2 shows that this system is also eclipsing, on a period of $205.965 \pm 0.014$\,min, but here the eclipses are V-shaped and only 0.5\,mag deep. Its low emission-line strengths, orbital period and V-shaped eclipse unambiguously mark it as a novalike object. WHT photometry of SDSS\,J005050.88$+$000912.6 and SDSS\,J210449.94$+$010545.8 yields periods of $80.3 \pm 2.2$ and $103.62 \pm 0.12$\,min, respectively. Photometry of the seventh and final system, SDSS\,J165658.12$+$212139.3, shows only flickering. Our results strengthen the conclusion that the faint magnitude limit of the SDSS spectroscopic database implies that the sample of CVs contained in it has quite different characteristics to previously studied samples of these objects. Five of the six orbital periods measured here are shorter than the observed 2--3\,hr CV period gap. Two systems have periods very close to the minimum orbital period for hydrogen-rich CVs.
\end{abstract}



\section{Introduction}                                                                       \label{sec:intro}

Cataclysmic variables (CVs) are interacting binary stars containing a white dwarf primary star in a close orbit with a low-mass secondary star which fills its Roche lobe. In most of these systems the secondary component is unevolved and is transferring material to the white dwarf primary via an accretion disc. Comprehensive reviews of the properties of CVs have been given by \citet{Warner95book} and \citet{Hellier01book}.

The evolution of CVs is governed primarily by angular momentum loss (AML) from the binary orbit. This is important during the phase of common-envelope evolution, where an initially wide orbit is dramatically shrunk due to the ejection of the envelope. The efficiency of this process is poorly understood, although recent advances \citep{NelemansTout05mn,Beer+07mn} are promising (see \citealt{Me++07mn} for further discussion). For many objects this results in a close but detached binary system.

Further AML causes the binary to become semidetached and undergo mass transfer. This AML is thought to occur mainly in two ways: by the emissison of gravitational radiation \citep{Paczynski67aca} and due to magnetic braking via the stellar wind \citep{Schatzman62anap} which acts on the secondary star \citep{VerbuntZwaan81aa,Rappaport++82apj}. According to the standard theory of CV evolution, magnetic braking is the dominant AML mechanism for the longer-period systems but abruptly terminates when the secondary star becomes completely convective. This disruption occurs when the orbital period has decreased to approximately three hours \citep{SpruitRitter83aa}. The distended secondary star is then able to shrink until it returns to thermal equilibrium, so no longer fills its Roche lobe. This disrupted magnetic braking scenario was formulated to explain the 2--3\,hr `gap' in the orbital period distribution of CVs \citep{WhyteEggleton80mn,Knigge06mn}, which contains far fewer systems than which occur at adjacent periods.

Whilst the CV is in the period gap, AML continues more slowly via gravitational radiation until a period of about two hours, at which point the Roche lobe of the secondary shrinks onto the star's surface and mass transfer recommences. AML continues to shorten the orbital period until a minimum is reached at about 76\,min \citep{Knigge06mn}, where structural changes in the secondary star mean that further AML causes the orbital period to increase again. This results in faint `period bounce' systems with very low mass transfer rates \citep{Patterson98pasp}.


Theoretical studies of the evolution of CVs \citep{Dekool92aa,DekoolRitter93aa,Kolb93aa,KolbDekool93aa,Politano96apj,KolbBaraffe99mn} are able to reproduce some of the features of the observed orbital period distribution of CVs \citep{Downes+01pasp,RitterKolb03aa}, but there are several discrepancies between the predicted and observed distributions \citep{Gansicke05aspc}. Firstly, it is predicted that the vast majority of CVs have orbital periods shorter than 3\,hr, whereas in the observed sample the numbers of systems above and below this value are roughly equal. Secondly, the predicted minimum period is around 65\,min and there should be a large number of systems at this point due to the very slow rate of change of period with time. The minimum period in the observed CV sample is at approximately 76\,min (excluding a small number of systems which contain evolved secondary stars) and the observational sample does not contain a `spike' at this period. Thirdly, population synthesis studies consistently find that at the present age of the Galaxy about 70\% of CVs should have evolved beyond the minimum period. Whilst there are over six hundred known CVs \citep{RitterKolb03aa} there are very few candidate period bouncers \citep[e.g.][]{Beuermann+00aa,Littlefair++03mn,Araujo+05aa,Patterson++05pasp}, and there are only {\em two} systems with a measured secondary star mass which is definitely below the hydrogen-burning limit: the eclipsing CVs SDSS\,J103533.02+055158.3 \citep{Littlefair+06sci,Me+06mn} and SDSS\,J150722.30+523039.8 \citep{Littlefair+07}.

A further problem with the standard disrupted magnetic braking evolutionary scenario is that there are many observations of very-low-mass fully convective stars which have strong magnetic fields \citep{Donati+06sci,Hallinan+06apj}, and no detections of a sudden cessation of magnetic braking in M\,dwarfs below a certain mass \citep{Andronov++03apj}.

\subsection{CVs identified by the Sloan Digital Sky Survey}

The known population of CVs is extremely heterogeneous due to the number of different ways in which individual objects have been discovered \citep{Gansicke05aspc}. Most of the methods of discovery are strongly biased towards detecting longer-period systems with higher mass transfer rates, because these are on average intrinsically brighter and bluer than short-period CVs. These effects mean that when comparing the observed population of CVs to theoretical predictions it is currently impossible to determine whether any discrepancies are due to the predictions being incorrect or to the large selection biases affecting the observed sample. The selection effects affecting observational studies of CVs have been considered in detail by \citet{Pretorius++07mn} for the case of the Palomar-Green survey \citep{Green++86apjs}. The selection biases in the known population of CVs are so strong and complex that this problem can only be reliably solved by studying a sample of CVs which has been homogeneously selected and is not strongly biased towards intrinsically bright or variable systems.

Recent studies of an homogeneous sample of CVs have included the population of objects spectroscopically identified in the Hamburg Quasar Survey \citep[HQS; ][]{Hagen+95aas}, which is a complete magnitude-limited sample conducted using objective prism observations. \citet{Aungwerojwit+05aa,Aungwerojwit+06aa} have found that the HQS CVs are on average photometrically less variable than the overall sample of known CVs, confirming that spectroscopic identification is important for avoiding some selection biases. These authors also found an unexpected deficiency of short-period dwarf novae in their sample, resulting in a low number ratio of short-period to long-period systems for the newly-discovered HQS CVs. The corresponding space density of short-period HQS CVs is much lower than theoretically predicted \citep[e.g.][]{Dekool92aa,Politano96apj}. Another sample arises from the 2dF Quasar Survey, which has serendipitously identified a number of CVs via their spectroscopic characteristics. \citet{Marsh+02conf} studied six of these objects and presented evidence that four of them had short periods, but the size of this sample must be increased before any detailed conclusions can be drawn.

\begin{table*} \begin{center}
\caption{\label{tab:iddata} Apparent magnitudes of our targets in the SDSS $ugriz$
passbands. \reff{$r_{\rm spec}$ is an apparent magnitude we have calculated by convolving
the SDSS flux-calibrated spectra with the $r$ passband function. It is obtained at a
different epoch to the $ugriz$ magnitudes measured from the CCD imaging observations,
but is less reliable as it is affected by slit losses, and any errors in astrometry
or positioning of the spectroscopic fibre entrance}.}
\begin{tabular}{lllccccccc} \hline
SDSS name                  & Short name  & Reference           & $u$ & $g$ & $r$ & $i$ & $z$ &  $r_{\rm spec}$ \\
\hline
SDSS J005050.88$+$000912.6 & SDSS\,J0050 & \citet{Szkody+05aj} & 20.37 & 20.37 & 20.34 & 20.48 & 20.89 & 20.68 \\
SDSS J013132.39$-$090122.2 & SDSS\,J0131 & \citet{Szkody+03aj} & 18.32 & 18.46 & 18.41 & 18.59 & 18.46 & 18.45 \\
SDSS J075443.01$+$500729.2 & SDSS\,J0754 & \citet{Szkody+06aj} & 17.38 & 17.24 & 17.24 & 17.29 & 17.31 & 17.47 \\
SDSS J155531.99$-$001055.0 & SDSS\,J1555 & \citet{Szkody+02aj} & 19.08 & 19.34 & 19.11 & 19.07 & 18.78 & 19.22 \\
SDSS J165658.12$+$212139.3 & SDSS\,J1656 & \citet{Szkody+05aj} & 18.01 & 18.51 & 18.28 & 18.18 & 18.05 & 18.28 \\
SDSS J205914.87$-$061220.4 & SDSS\,J2059 & \citet{Szkody+02aj} & 18.20 & 18.38 & 18.40 & 18.40 & 18.40 & 18.81 \\
SDSS J210449.94$+$010545.8 & SDSS\,J2104 & \citet{Szkody+06aj} & 17.34 & 17.23 & 17.18 & 17.18 & 17.17 & 20.11 \\
\hline \end{tabular} \end{center} \end{table*}

\begin{table*} \begin{center}
\caption{\label{tab:obslog} Log of the observations presented in this work.}
\begin{tabular}{lcccccrrr} \hline
Target & Start date & Start time & End time & Telescope and & Optical  &  Number of   & Exposure \\
       & (UT)       &  (UT)      &  (UT)    &  instrument   & element  & observations & time (s) \\
\hline
SDSS\,J0050 & 2006 09 28 & 01 59 & 05 21 & WHT Aux Port & SDSS $g$ filter &  167 &  60 \\
SDSS\,J0050 & 2006 09 28 & 01 19 & 05 37 & WHT Aux Port & SDSS $g$ filter &  188 &  60 \\[3pt]
SDSS\,J0131 & 2006 09 26 & 01 59 & 05 04 & WHT ISIS     & R600B R316R     &   23 & 600 \\
SDSS\,J0131 & 2006 09 26 & 23 40 & 02 13 & WHT ISIS     & R600B R316R     &   15 & 600 \\[3pt]
SDSS\,J0754 & 2007 01 23 & 20 17 & 01 22 & NOT ALFOSC   & (unfiltered)    & 1127 &  15 \\
SDSS\,J0754 & 2007 02 09 & 16 26 & 23 54 & Tuorla SBIG  & (unfiltered)    &  150 & 240 \\
SDSS\,J0754 & 2007 02 09 & 21 06 & 01 20 & Tuorla SBIG  & (unfiltered)    &  102 & 240 \\[3pt]
SDSS\,J1555 & 2006 05 18 & 00 17 & 05 30 & INT WFC      & (unfiltered)    &  183 &  50 \\
SDSS\,J1555 & 2006 05 18 & 23 57 & 01 13 & INT WFC      & (unfiltered)    &   50 &  50 \\[3pt]
SDSS\,J1656 & 2006 05 16 & 01 24 & 05 33 & INT WFC      & (unfiltered)    &  171 &  40 \\[3pt]
SDSS\,J2059 & 2006 08 23 & 00 08 & 03 06 & NOT ALFOSC   & (unfiltered)    &  553 &  15 \\
SDSS\,J2059 & 2006 09 24 & 20 10 & 00 20 & WHT ISIS     & R600B R316R     &   24 & 600 \\
SDSS\,J2059 & 2006 09 25 & 20 27 & 22 46 & WHT ISIS     & R600B R316R     &   14 & 600 \\[3pt]
SDSS\,J2104 & 2006 09 27 & 20 02 & 22 55 & WHT Aux Port & SDSS $g$ filter &  145 &  60 \\
SDSS\,J2104 & 2006 09 28 & 00 10 & 01 55 & WHT Aux Port & SDSS $g$ filter &   88 &  60 \\
SDSS\,J2104 & 2006 09 28 & 20 04 & 23 23 & WHT Aux Port & SDSS $g$ filter &  142 &  60 \\
\hline \end{tabular} \end{center} \end{table*}

We are conducting a research program to measure the orbital periods of objects identified as CVs from spectra obtained by the Sloan Digital Sky Survey (SDSS\footnote{\tt http://www.sdss.org/}; \citealt{York+00aj}). The selection biases in this sample of objects are expected to be much less severe than those in other CV samples (see below), so they should be much more representative of the intrinsic population of CVs than any other known sample of these objects. To date, a total of 212 CVs have been discovered by the SDSS \citep{Szkody+02aj, Szkody+03aj, Szkody+04aj, Szkody+05aj, Szkody+06aj, Szkody+07aj}, and more are expected to be found from future observations. About 90 of the SDSS CVs have measured periods, of which half have been obtained by our group (\citealt{Gansicke+06mn}; \citealt{Me+06mn}, hereafter Paper\,I; \citealt{Me+07mn}; Dillon et al.\ 2007, in preparation).

The SDSS has obtained spectroscopy of a large number of point sources whose $ugriz$ colour indices were of interest to one of the automatic targeting algorithms (particularly for quasars); the resulting set of CVs has been identified on the basis of their spectral characteristics. This sample is therefore not strongly biased towards blue objects, those with high mass transfer rates, or systems which have significant X-ray emission. It is a magnitude-limited sample, but goes to much fainter apparent magnitudes ($g \sim 21$) than other similar surveys so this bias should be comparatively weak. There is a bias resulting from the colour-dependent target selection, but this is also minor as the SDSS spectroscopic database covers a wide volume of colour space. As the SDSS main survey points out of the Galactic plane, and its spectroscopic observations have a faint limiting magnitude, the resulting sample of CVs is approximately volume-limited for the intrinsically brighter longer-period objects (see Section\,\ref{sec:discussion}).

In this work we present time-resolved spectroscopy and photometry of seven CVs discovered by the SDSS (Table\,\ref{tab:iddata}) and obtain orbital periods for six of these, two of which are eclipsing. Five of the six system have short orbital periods ($\Porb \la 2$\,hr). In this work we shall abbreviate the names of the targets to SDSS\,J0050, SDSS\,J0131, SDSS\,J0754, SDSS\,J1555, SDSS\,J1656, SDSS\,J2059 and SDSS\,J2104. Their full names and $ugriz$ apparent magnitudes are given in Table\,\ref{tab:iddata}.


\section{Observations and data reduction}\label{sec:obs}

\subsection{WHT spectroscopy}                                                       \label{sec:obs:whtspec}

\begin{figure*}
\includegraphics[width=\textwidth,angle=0]{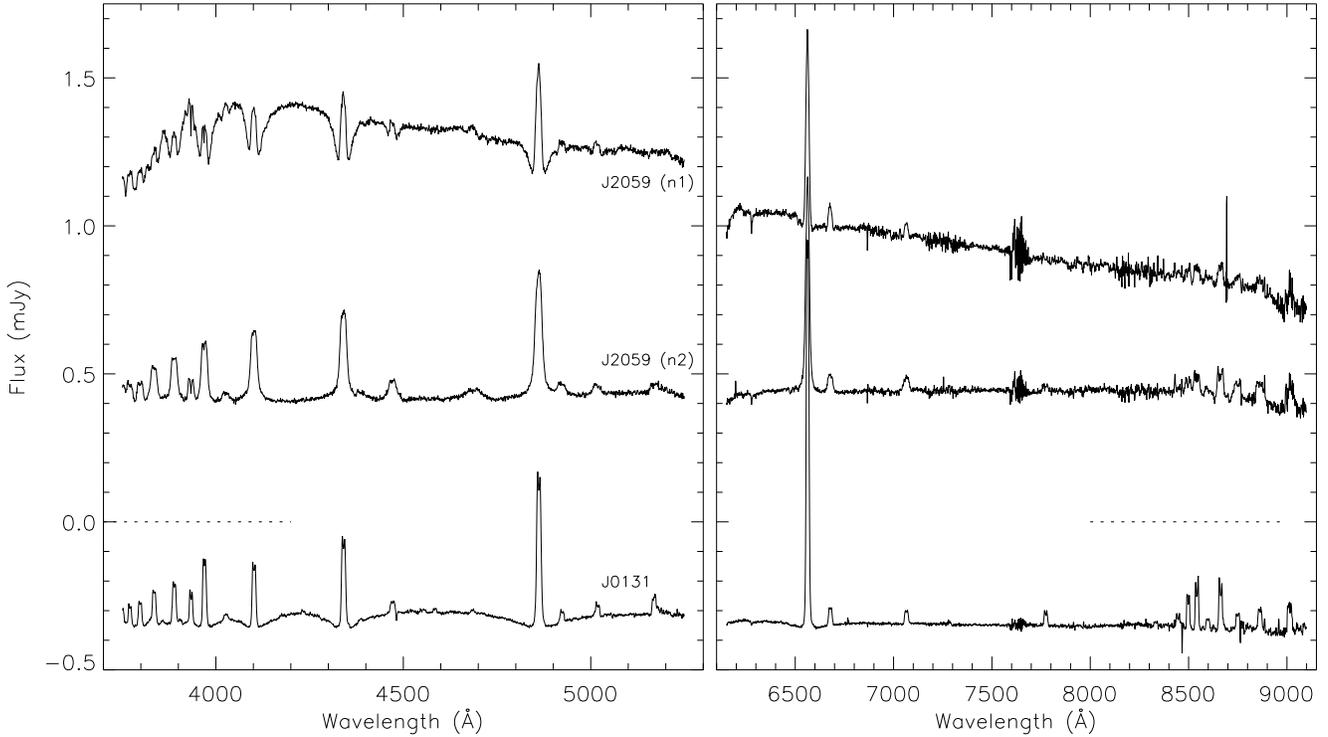}
\caption{\label{fig:whtspec} Flux-calibrated averaged WHT/ISIS spectra
of the CVs SDSS\,J0131 and SDSS\,J2059. The spectra for SDSS\,J0131 are
offset by $-$0.5\,mJy for clarity. For SDSS\,J2059 the spectra from our
two nights of observations have been averaged separately to illustrate
the change in its spectral characteristics as it was declining from
outburst. Data from the spectrograph blue arm are shown in the left panel,
and from the red arm in the right panel.} \end{figure*}

\begin{figure}
\includegraphics[width=0.23\textwidth,angle=0]{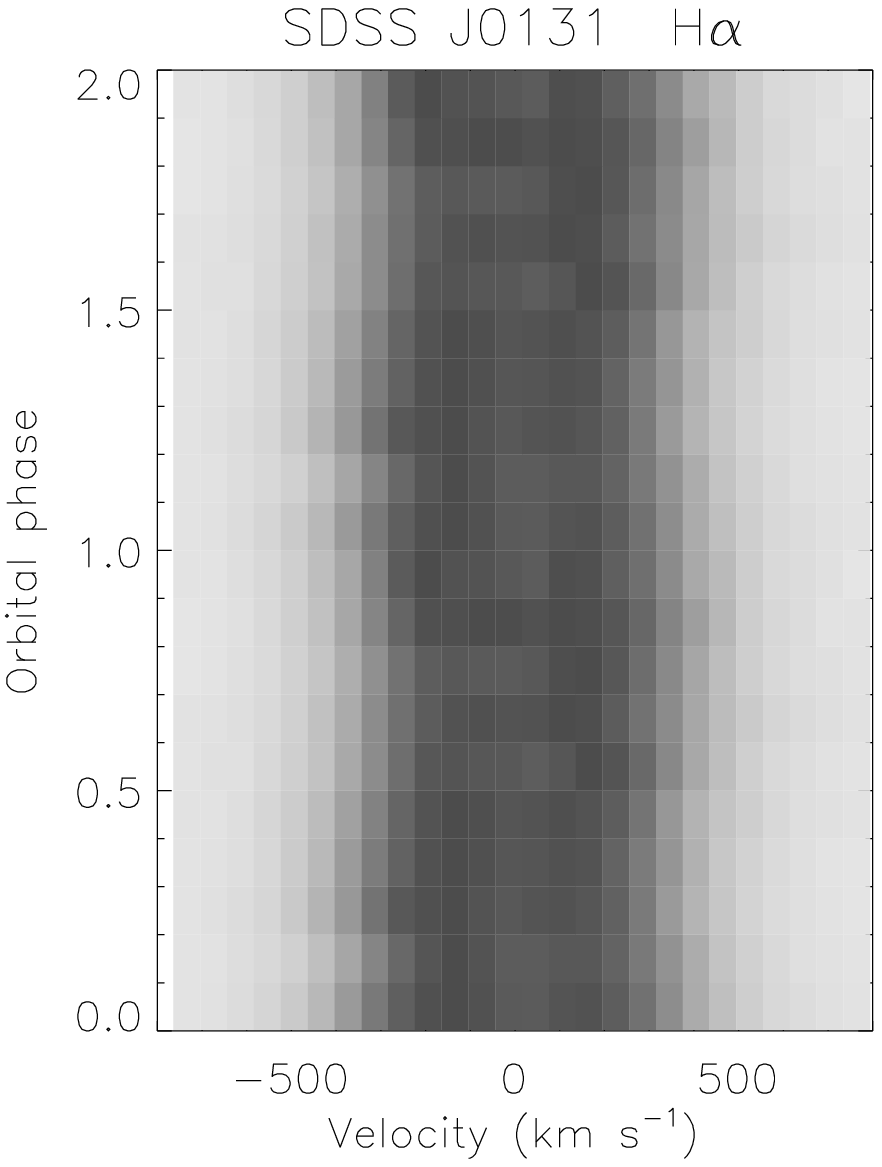}
\includegraphics[width=0.23\textwidth,angle=0]{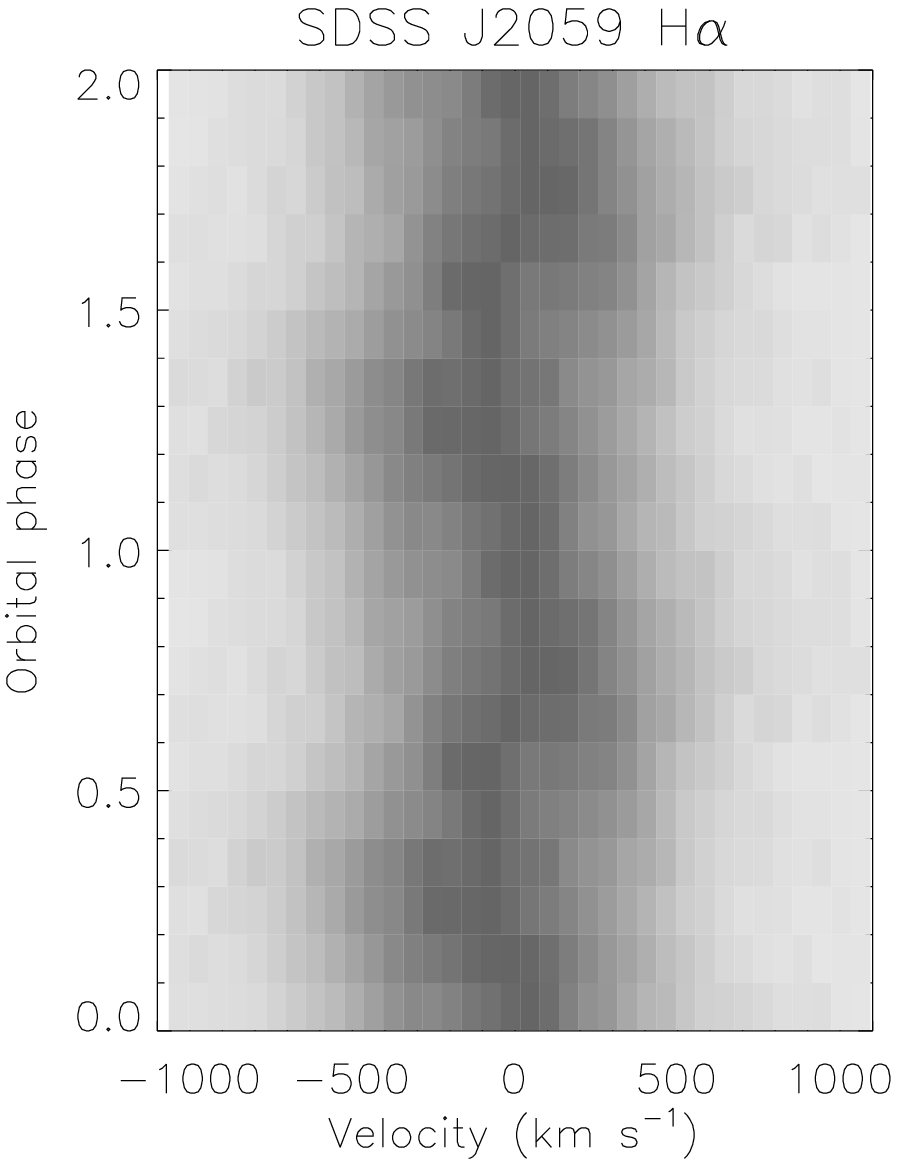}
\caption{\label{fig:trailed} Greyscale plots of the continuum-normalised
and phase-binned trailed spectra of SDSS\,J0131 and SDSS\,J2059. In each
case the H$\alpha$ line is shown as it gives the strongest signal.}
\end{figure}

Spectroscopic observations of SDSS\,J0131 and SDSS\,J2059 were obtained in 2006 September using the ISIS double-beam spectrograph on the William Herschel Telescope (WHT) at La Palma (Table\,\ref{tab:obslog}). For the red arm we used the R316R grating and Marconi CCD binned by factors of 2 (spectral) and 4 (spatial), giving a wavelength range of 6300--9200\,\AA\ and a reciprocal dispersion of 1.66\,\AA\ per binned pixel. For the blue arm we used the R600B grating and EEV12 CCD with the same binning factors as for the Marconi CCD, giving a wavelength coverage of 3640--5270\,\AA\ at 0.88\,\AA\ per binned pixel. From measurements of the full widths at half maximum (FWHMs) of arc-lamp and night-sky spectral emission lines, we find that this gave resolutions of 3.5\,\AA\ (red arm) and 1.8\,\AA\ (blue arm).


\begin{figure} \includegraphics[width=0.48\textwidth,angle=0]{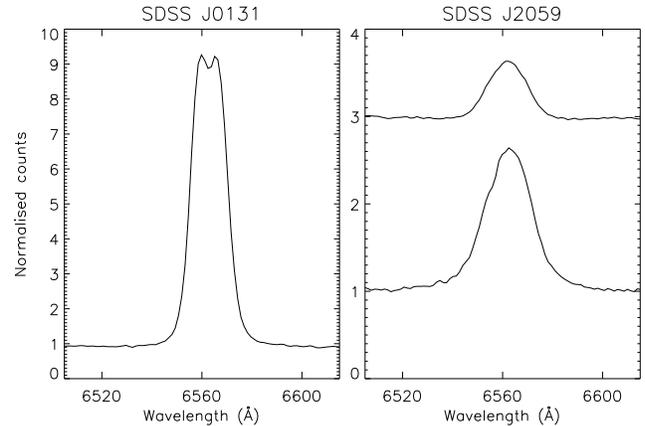} \\
\caption{\label{fig:Halpha} The averaged and continuum-normalised H$\alpha$
emission line profiles of the CVs for which we obtained WHT/ISIS spectra.
For SDSS\,J2059 we plot the averaged spectra from the two nights separately as
it was declining from outburst during our observations. The first night is
offset by two from the second night.} \end{figure}

Data reduction was undertaken using optimal extraction (\citealt{Horne86pasp}) as implemented in the {\sc pamela}\footnote{{\sc pamela} and {\sc molly} were written by TRM and can be found at {\tt http://www.warwick.ac.uk/go/trmarsh}} code \citep{Marsh89pasp}, which also makes use of the {\sc starlink}\footnote{The Starlink Software Group homepage can be found at {\tt http://www.starlink.rl.ac.uk/}} packages {\sc figaro} and {\sc kappa}. Copper-neon and copper-argon arc lamp exposures were taken every hour during observations and the wavelength calibration for each science exposure was interpolated from the two arc lamp observations which bracketed it. We removed the telluric lines and flux-calibrated the target spectra using observations of G191-B2B, treating each night separately.

\begin{figure} \includegraphics[width=0.48\textwidth,angle=0]{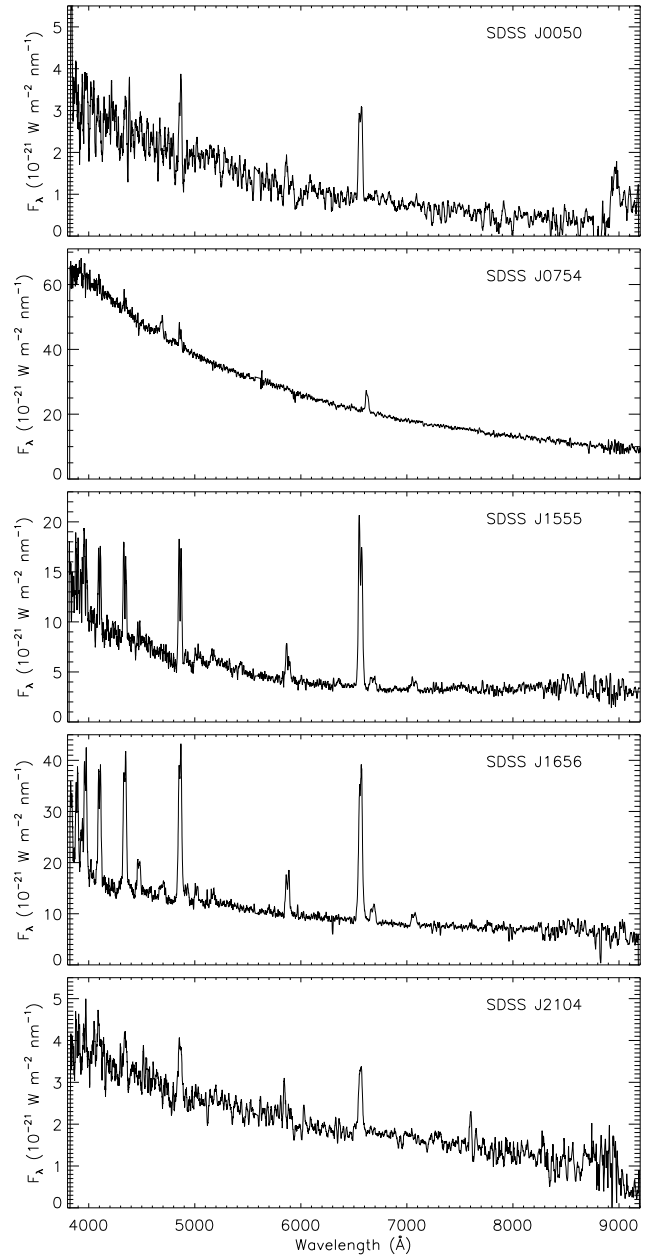} \\
\caption{\label{fig:sdssspec} SDSS spectra of the four CVs for which we did not
obtain WHT/ISIS observations. For this plot the flux levels have been smoothed
with 10- or 18-pixel Savitsky-Golay filters using the {\sc savgol} procedure
in {\sc IDL} (see {\tt http://www.rsinc.com/}). The units of the abscissae are
$10^{-21}$\,W\,m$^{-2}$\,nm$^{-1}$, which corresponds to
$10^{-17}$\,erg\,s$^{-1}$\,cm$^{-2}$\,\AA$^{-1}$.}\end{figure}

The averaged WHT spectra are shown in Fig.\,\ref{fig:whtspec}. Trailed greyscale plots of the phase-binned spectra are given in Fig.\,\ref{fig:trailed} and close-ups of the H$\alpha$ profiles are shown Fig.\,\ref{fig:Halpha}. For reference, the SDSS spectra of the objects without WHT spectroscopy are plotted in Fig.\,\ref{fig:sdssspec}.

\subsection{WHT, INT and NOT photometry}                                               \label{sec:obs:phot}

Photometric observations were obtained for SDSS\,J0050 and SDSS\,J2104 in 2006 September (Table\,\ref{tab:obslog}) using the WHT Auxiliary Port Imaging Camera and Tek CCD binned by a factor of 2 in both spatial directions to give a plate scale of 0.22\as\,px$^{-1}$. The SDSS $g$ filter was used and exposure times for these faint objects were 60\,s. \reff{The amount of signal in the spectrum of SDSS\,J2104 (Fig.\ref{fig:sdssspec}) is much smaller than expected from its $ugriz$ magnitudes from the SDSS imaging observations (Table\,\ref{tab:iddata}). For this reason we expected the CV to be too faint to be observed spectroscopically with the WHT, so studied it photometrically instead (see Section\,\ref{sec:2104}).}

We obtained differential photometry of SDSS\,J0754 and SDSS\,J2059 using the Nordic Optical Telescope (NOT) and ALFOSC imager in 2006 August and 2007 January. The observations were unfiltered, and a high cadence was achieved through the use of short exposure times (15\,s), binning by a factor of 2 in both spatial directions, and heavy windowing. The detector was an EEV 2k$\times$4k pixel CCD with an unbinned plate scale of 0.19\as\,px$^{-1}$.

We obtained unfiltered differential photometry of SDSS\,J0754 on two nights in 2007 February using the 0.7\,m Schmidt-V\"ais\"al\"a telescope and SBIG ST-8 CCD camera which was windowed to reduce readout time.

Unfiltered differential photometry of SDSS\,J1555 and SDSS\,J1656 was obtained in 2006 May using the Isaac Newton Telescope (INT) at La Palma. The INT was equipped with the Wide Field Camera (WFC), an array of four EEV 2k$\times$4k pixel CCDs with a plate scale of 0.37\as\,px$^{-1}$, and no binning was used.

All photometric data were reduced using the pipeline described by \citet{gansicke+04aa}, which performs bias and flat-field corrections within {\sc midas}\footnote{\tt http://www.eso.org/projects/esomidas/} and uses the {\sc sextractor} package \citep{BertinArnouts96aas} to perform aperture photometry for all objects in the field of view.

Instrumental differential magnitudes were converted into apparent $g$ (or approximately $g$) magnitudes using several comparison stars near the target along with their brightnesses given on the SDSS website.


\section{Data analysis}

\begin{table*} \begin{center}
\caption{\label{tab:orbits} Circular spectroscopic orbits for SDSS\,J0131 and
SDSS\,J2059 found using {\sc sbop}. The adopted orbits are shown in bold type.
Orbits fitted using periods which are one-day aliases of the adopted value are
given in normal type for comparison with the adopted orbits. $\sigma_{\rm rms}$
is the root mean square of the residuals of the observations around the best
fit. The reference times (corresponding to zero phase) refer to inferior
conjuction (the midpoint of primary eclipse in eclipsing systems).}
\begin{tabular}{l r@{\,$\pm$\,}l r@{\,$\pm$\,}l r@{\,$\pm$\,}l r@{\,$\pm$\,}l r}\hline
Target&\mc{Orbital period}&\mc{Reference time}&\mc{Velocity amplitude}&\mc{Systemic velocity}&$\sigma_{\rm rms}$\\
      &     \mc{(day)}    &    \mc{(HJD)}     &       \mc{(\kms)}     &      \mc{(\kms)}     &      (\kms)      \\
\hline 
SDSS\,J0131  &  0.053191 & 0.00011  &  54000.3172 & 0.0101  & \ \ \ \ \ 18.3 & 1.8  &     8.4 & 1.4  &  8.1 \\
SDSS\,J0131  &  {\bf 0.056627} & {\bf 0.00009}  &  {\bf 54000.3189} & {\bf 0.0076}  &           {\bf 19.6} & {\bf 1.4}  &     {\bf 7.7} & {\bf 1.0}  &  {\bf 6.2} \\
SDSS\,J0131  &  0.605348 & 0.00012  &  54000.3211 & 0.0096  &           19.0 & 1.7  &     6.9 & 1.2  &  7.2 \\[3pt]
SDSS\,J2059  &  0.069435 & 0.00012  &  54000.0920 & 0.0061  &           57.2 & 4.1  & $-$36.5 & 2.9  & 17.6 \\
SDSS\,J2059  &  {\bf 0.074665} & {\bf 0.00010}  &  {\bf 54000.0654} & {\bf 0.0050}  &           {\bf 58.5} & {\bf 3.0}  & {\bf $-$36.8} & {\bf 2.2}  & {\bf 13.3} \\
SDSS\,J2059  &  0.080733 & 0.00016  &  54000.0349 & 0.0076  &           56.1 & 4.1  & $-$37.2 & 2.9  & 17.8 \\
\hline \end{tabular} \end{center} \end{table*}

\begin{table} \begin{center}
\caption{\label{tab:boot} Results of the bootstrapping simulations
(see Section\,\ref{sec:data:period} for details). They are given as a
percentage of the total number of simulations returning a periodogram
with the frequency of the highest power within the peak of the specific
alias in the original periodogram. These can be directly interpreted as
the probability that a specific alias is the correct one. However, we
expect these results to be conservative for several reasons
(Section\,\ref{sec:data:period}). Frequencies are given in \cd. The
results for SDSS\,J0050 are given for {\em double} the orbital frequency
because the light curve is double-humped and the periodogram at the
orbital frequency has a more complex variation. In a few cases the
probabilities for one set of simulations do not add up to 100\% because
a few points fell outside the range of the frequencies included in this
table.}
\begin{tabular}{lcccccc} \hline
SDSS\,J0050       & \multicolumn{6}{c}{Frequency (\cd)}     \\
                  & 34.0 & 35.0 & 35.9 & 36.9 & 37.9 & 38.9 \\
      \hline
Scargle           &  0.0 &  0.4 & 13.8 & 52.3 & 31.6 &  1.5 \\
AoV               &  0.0 &  4.6 & 25.9 & 35.1 & 33.1 &  1.1 \\
ORT               &  0.7 &  4.7 & 18.8 & 48.7 & 24.3 &  2.7 \\
\hline \\ \hline
SDSS\,J0131       & \multicolumn{6}{c}{Frequency (\cd)}     \\
                  & 14.3 & 15.5 & 16.6 & 17.7 & 18.8 & 20.0 \\
      \hline
Scargle           &      &  0.9 & 14.6 & 83.8 &  0.7 &      \\
AoV               &  0.4 &  1.1 &  4.7 & 90.8 &  2.7 &  0.3 \\
ORT               &  1.3 &  2.0 & 16.6 & 76.9 &  3.2 &      \\
\hline \\ \hline
SDSS\,J2059       & \multicolumn{6}{c}{Frequency (\cd)}     \\
                  & 10.4 & 11.4 & 12.4 & 13.4 & 14.4 & 15.5 \\
      \hline
Scargle           &      &      &  0.1 & 98.4 &  1.5 &      \\
AoV               &      &      &  2.4 & 77.3 & 20.3 &      \\
ORT               &      &      &  2.1 & 95.4 &  2.5 &      \\
\hline \\ \hline
SDSS\,J2104       & \multicolumn{6}{c}{Frequency (\cd)}     \\
                  & 11.8 & 12.9 & 13.9 & 14.9 & 15.9 &      \\
      \hline
Scargle           &      &      & 99.8 &  0.2 &      &      \\
AoV               &      &      &100.0 &      &      &      \\
ORT               &      &  0.1 & 99.9 &      &      &      \\
\hline \end{tabular} \end{center} \end{table}

\subsection{Radial velocity measurement}                                                  \label{sec:data:rv}

We measured radial velocities from emission lines in the spectra of our targets by cross-correlation with single and double Gaussian functions \citep{SchneiderYoung80apj}, as implemented in {\sc molly}\footnote{The reduced spectra and radial velocity observations presented in this work will be made available at the CDS ({\tt http://cdsweb.u-strasbg.fr/}) and at {\tt http://www.astro.keele.ac.uk/$\sim$jkt/}}. In each case we tried a range of different widths and separations for the Gaussians in order to verify the consistency of our results (see Paper\,I for further details).

\subsection{Orbital period measurement}                                             \label{sec:data:period}

The measured radial velocities and differential magnitudes for each CV were searched for periods using periodograms computed using the \citet{Scargle82apj} method, analysis of variance \citep[AoV; ][]{Schwarzenberg89mn} and orthogonal polynomials \citep[ORT; ][]{Schwarzenberg96apj}, as implemented within the {\sc tsa}\footnote{\scriptsize\tt http://www.eso.org/projects/esomidas/doc/user/98NOV/volb/node220.html} context in {\sc midas}. In general two Fourier terms were used for ORT, which is appropriate for the relatively simple variation in light or radial velocity for these objects.

To find the final values of the orbital periods, and to investigate the aliases in the periodograms, we fitted circular spectroscopic orbits (sine curves) to the data using the {\sc sbop}\footnote{Spectroscopic Binary Orbit Program, written by P.\ B.\ Etzel, \\ {\tt http://mintaka.sdsu.edu/faculty/etzel/}} program, which we have previously found to give reliable error estimates for the optimised parameters \citep{Me+05mn}. Further details of our methods are given in Paper\,I. The parameters of the final spectroscopic orbits, as well as the individual orbital fits for the different possible aliases, are given in Table\,\ref{tab:orbits}.

To assess the likelihood of the point of highest power corresponding to the actual orbital period we have performed bootstrapping simulations (see Paper\,I) by randomly resampling the data with replacement and calculating a new periodogram a total of 1000 times \citep[][p.\,686]{Press+92book}. The fraction of periodograms in which the highest peak fell close to a particular alias can be interpreted as the likelihood of that alias being correct. However, in these cases we expect the resulting probabilities for the correct peak to be quite conservative (i.e.\ too low) for two reasons. Firstly, because the simulated datasets necessarily contain fewer unique epochs than the original data some temporal definition is sacrificed. The bootstrapping periodograms are always clearly inferior to that calculated from the actual data, confirming this picture. Secondly, when picking the best alias interactively we use more information than just the periodogram (including the shape of the phased radial velocity curve). The results from the bootstrapping analysis are given in Table\,\ref{tab:boot}.

\section{Results for each system}

\subsection{SDSS J005050.88$+$000912.6}                                                       \label{sec:0050}

\begin{figure} \includegraphics[width=0.48\textwidth,angle=0]{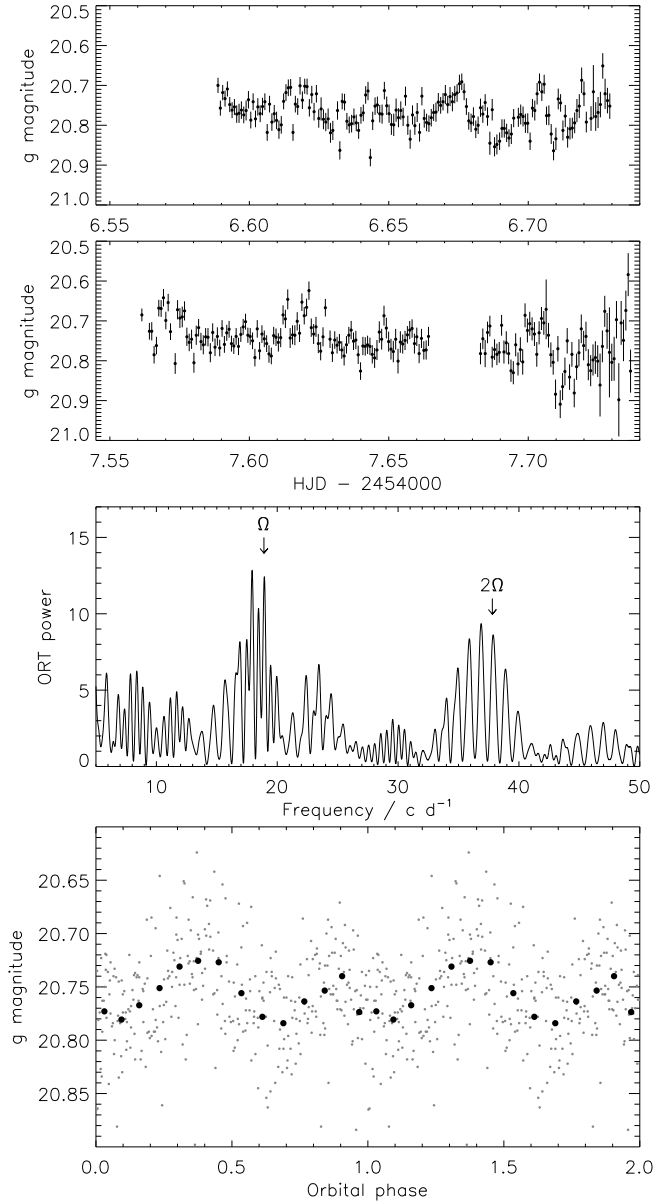}\\
\caption{\label{fig:0050:lcplot} WHT imaging photometry of SDSS\,J0050. The
data were obtained using an SDSS $g$ filter. Data from the individual nights
are shown in the top two panels. The two-harmonic ORT periodogram of the data
is given in the third panel, with the orbital frequency ($\Omega$) and second
harmoninc indicated. The lowest panel shows the phased data (grey points) and
these data combined into 15 phase bins (black points). Note that the magnitude
calibration is quite approximate as the only available comparison star is quite
red so has very different colours to SDSS\,J0050.} \end{figure}

SDSS\,J0050 was discovered to be a CV by \citet{Szkody+05aj} from its SDSS spectrum, which shows weak double-peaked emission at H$\alpha$ and H$\beta$. This spectrum is very noisy as the object is faint (Fig.\,\ref{fig:sdssspec}), but strong smoothing reveals the presence of wide absorption at H$\beta$ arising from the white dwarf. This indicates that the accretion disc is dim and that the system as a whole is intrinsically faint.

We obtained eight hours of photometry with the WHT Auxiliary port, spread over two consecutive nights (Fig.\,\ref{fig:0050:lcplot}). Periodograms of these data show significant power at around 40\,min and 80\,min. We interpret this as a double-humped light curve arising from a CV with a period close to the 76\,min minimum value \citep{Knigge06mn}. Whilst the periodograms have a substantial number of possible aliases at periods of 75 to 90\,min, bootstrapping simulations show that only three periods in the region of 40\,min give a good signal. These three periods all have corresponding signals at double their periods in the periodograms. Assuming that the signals around 40\,min are at double the orbital period, sine curve fits give possible orbital periods of $76.03 \pm 0.10$\,min, $78.10 \pm 0.10$\,min and $80.28 \pm 0.12$\,min. The phased light curves have the least scatter for the last of these possibilities, and significantly more for the first, but we cannot securely pick the correct alias. We will therefore take the longest of the three periods to be correct but quote an uncertainty which includes the next best period: $80.3 \pm 2.2$\,min. The results of the bootstrapping simulations are given in Table\,\ref{tab:boot} for the frequencies close to double the orbital period.


\subsection{SDSS J013132.39$-$090122.2}                                                       \label{sec:0131}

\begin{figure} \includegraphics[width=0.48\textwidth,angle=0]{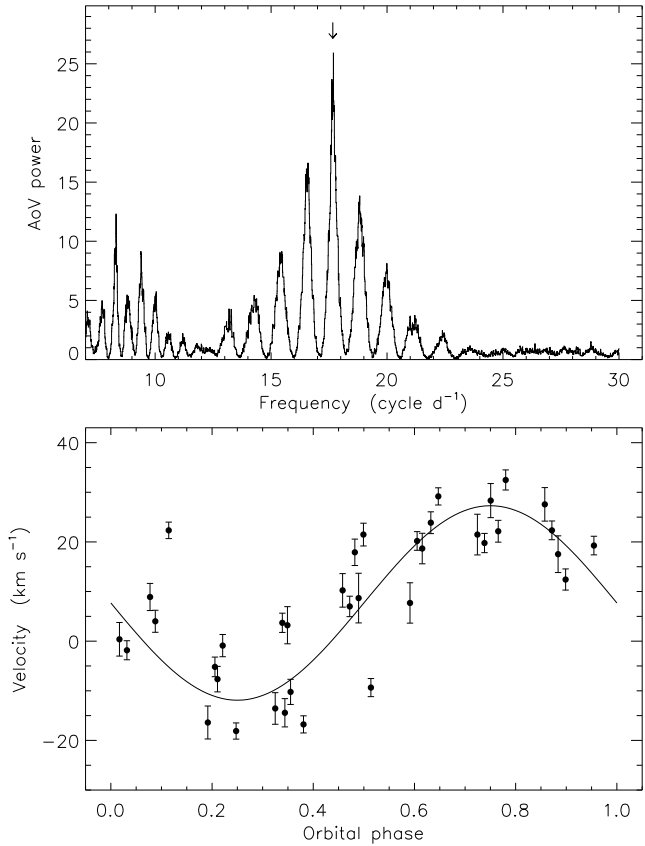}\\
\caption{\label{fig:0131:rvplot} AoV periodogram (top) and phased radial velocity
curve (bottom) for SDSS\,J0131. The radial velocities were measured from the
H$\alpha$ emission lines in our WHT/ISIS spectra using the double Gaussian
technique (see text for details). The frequency corresponding to the best-fitting
orbital period from {\sc sbop} is indicated with a downward-pointing arrow.} \end{figure}

SDSS\,J0131 was discovered to be a CV by \citet{Szkody+03aj} on the basis of its SDSS spectrum, which shows strong and slightly double-peaked hydrogen Balmer emission (Fig.\,\ref{fig:Halpha}) along with emission lines from \ion{He}{I}, \ion{Ca}{II} and \ion{Fe}{II} which are typical of dwarf novae. The infrared \ion{Ca}{II} triplet emission is similar to that seen in other CVs \citep{Persson98pasp} and also reminiscent of the emission from the accreting white dwarf SDSS\,J122859.93+104032.9 \citep{Gansicke+06sci}. Trailed spectra show no absorption or emission lines which could be attributed to the secondary star. With the exception of H$\alpha$, a wide absorption component is present in the Balmer lines from the white dwarf (Fig.\,\ref{fig:whtspec}), indicating that the accretion disc is faint and that the mass transfer rate is low. \citet{Szkody+03aj} found an orbital period of approximately 98\,min from two hours of follow-up spectroscopic observations.

Some white dwarfs in CVs with low mass transfer rates demonstrate ZZ\,Ceti non-radial pulsations, the prototype of this class being GW\,Lib \citep{WarnerVanzyl98iaus,Vanzyl+00balta,Vanzyl+04mn}. \citet{WarnerWoudt04asp} obtained high-speed photometry of SDSS\,J0131, finding that the primary component is a ZZ\,Ceti pulsator with frequencies of 1680 and 3030\,$\mu$Hz (575 and 300\,s). \citet{Szkody+07apj} obtained ultraviolet low-resolution HST spectra, finding a pulsation period of 212\,s. From fitting their observations with model white dwarf spectra \citep[see][]{Gansicke+05apj} they found a white dwarf temperature of $14\,500 \pm 1500$\,K and a distance of $360 \pm 100$\,pc for the system. This distance gives an absolute magnitude at the time of the SDSS imaging observations of $M_g = 10.7 \pm 0.6$.

We obtained a total of 38 WHT/ISIS spectra of SDSS\,J0131 over two nights. Radial velocities were measured using the double Gaussian technique for the H$\alpha$, H$\beta$ and H$\gamma$ emission lines, and searched for periods using Scargle, AoV and ORT periodograms. In every case the strongest power was at a period of 81.4\,min (Fig.\,\ref{fig:0131:rvplot}, Table\,\ref{tab:boot}). The best results were found using Gaussians of FWHM 300\kms\ and separation 1000\kms\ for H$\alpha$. For the H$\beta$ and H$\gamma$ emission lines, which contain less flux than H$\alpha$, the best results were found for smaller separations. This periodicity is also clearly present in radial velocities measured from the cores of the line.

We have fitted a sine curve to the H$\alpha$ radial velocities to find the best period (Fig.\,\ref{fig:0131:rvplot}). The one-day aliases give clearly inferior fits (Table\,\ref{tab:orbits}). After rejecting one clearly discrepant observation, we find a period of $81.54 \pm 0.13$\,min. Bootstrapping simulations using the AoV algorithm (Table\,\ref{tab:boot}) confirm that this period alias is 90\% likely to correspond to the correct period (but remember that we expect this figure to be quite conservative).

The spectroscopic orbit for SDSS\,J0131 has a low velocity amplitude of $K = 19.6 \pm 1.4$\kms\ (Table\,\ref{tab:orbits}). This amplitude is reminiscent of the $13.8 \pm 1.6$\kms\ value found for SDSS\,J121607.03+052013.9 (Paper\,I) and suggests that the mass ratio (which we define as $q = \frac{M_2}{M_{\rm WD}}$ where $M_{\rm WD}$ and $M_2$ are the primary and secondary star masses, respectively) of this CV is extremely low, and that the mass of the secondary star may lie in the brown dwarf regime. A trailed greyscale plot of the H$\alpha$ profile versus orbital phase (Fig.\,\ref{fig:trailed}) shows a slight orbital modulation but no evidence of emission from a bright spot. We note that \citet{Szkody+03aj} found $K = 29 \pm 5$\kms\ for H$\alpha$ and $K = 25 \pm 2$\kms\ for H$\beta$, values reasonably consistent with that determined here.

\subsubsection{Constraining the stellar physical properties of SDSS\,J0131}            \label{sec:0131:constraints}

\begin{figure} \includegraphics[width=0.48\textwidth,angle=0]{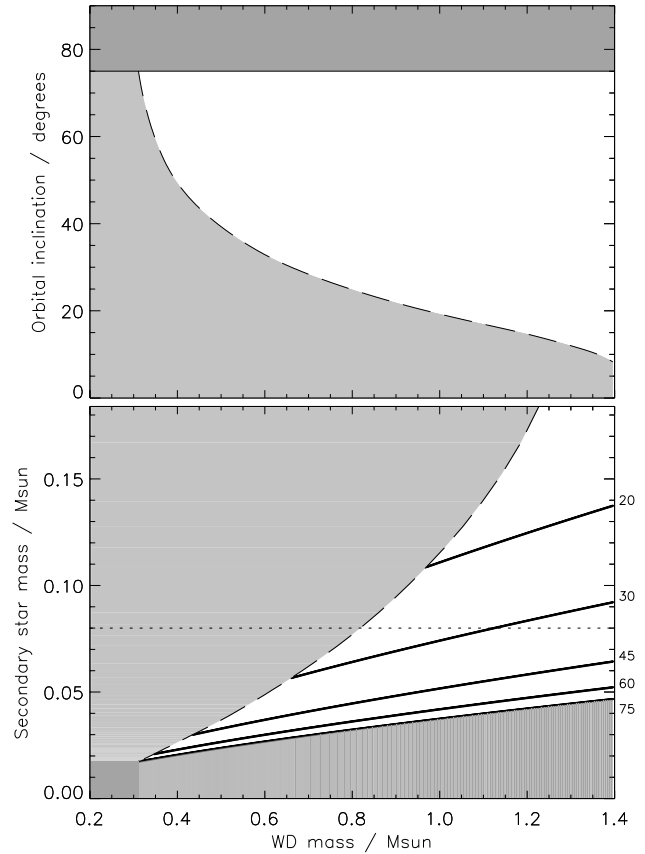} \\
\caption{\label{fig:0131:params} Diagram showing constraints on the possible properties
of SDSS\,J0131 (see Section\,\ref{sec:0131:constraints} for details). The dark shaded
regions indicate areas where the orbital inclination would be 75$^\circ$ or more, which
is ruled out due to the observed absence of eclipses. The light shaded regions indicate
parameter space which is eliminated as a result of the observed FWZI of the H$\alpha$
emission. This sets a lower limit on the radius of inner edge of the accretion disc,
which must be outside the surface of the white dwarf. The dashed lines indicate the
edge of this region. The solid lines indicate the locus of points satisfying the mass
function for orbital inclinations of 75$^\circ$, 60$^\circ$, 45$^\circ$, 30$^\circ$
and 20$^\circ$ (labelled). The canonical minimum mass for core hydrogen burning is
shown using a dotted line.} \end{figure}

As the spectroscopic orbit we have measured for SDSS\,J0131 has a small velocity amplitude, indicative of an extremely low mass ratio, we have attempted to constrain its properties in the same way as for the CV SDSS\,J121607.03+052013.9 (Paper\,I). The spectra show no sign of eclipses (which can be contrasted with SDSS\,J103533.02+055158.3; Paper\,I), indicating that the inclination is below about 75$^\circ$ (\citealt{Hellier01book}). This constraint is indicated on Fig.\,\ref{fig:0131:params} using darker-grey shading.

We have measured the full width at zero intensity (FWZI) of the mean H$\alpha$ emission line profile in our WHT spectra to be $70 \pm 10$\,\AA, corresponding to a half-width zero-intensity velocity of $1600 \pm 200$\kms. We have adopted a mass--radius relation for a 15\,000\,K helium-core white dwarf from \citet{Bergeron++95apj}, supplemented with the zero-temperature predictions of \citet{HamadaSalpeter61apj} for masses between 1.2\Msun\ and 1.4\Msun. Combining these two pieces of information gives the orbital inclination as a function of mass for which the innermost part of the disc is on a Keplerian orbit at the white dwarf's surface. This provides an upper limit on its radius which translates into a lower limit on its mass. This constraint is shown on Fig.\,\ref{fig:0131:params} by light grey shading in the areas of parameter space which are inaccessible.

The mass function of the orbit is $(4.4 \pm 1.0) \times 10^{-5}$\Msun. For a specified orbital inclination this gives the mass ratio of the stars (Fig.\,\ref{fig:0131:params}, right panel).

We have assumed that the orbital velocity amplitude found from the emission line wings is representative of the orbital movement of the white dwarf. It is generally thought that this overestimates the actual motion, which will translate directly into an overestimate of the mass of the secondary star (e.g.\ \citealt{Thoroughgood+05mn}; \citealt{Unda++06mn}).

\begin{figure} \includegraphics[width=0.48\textwidth,angle=0]{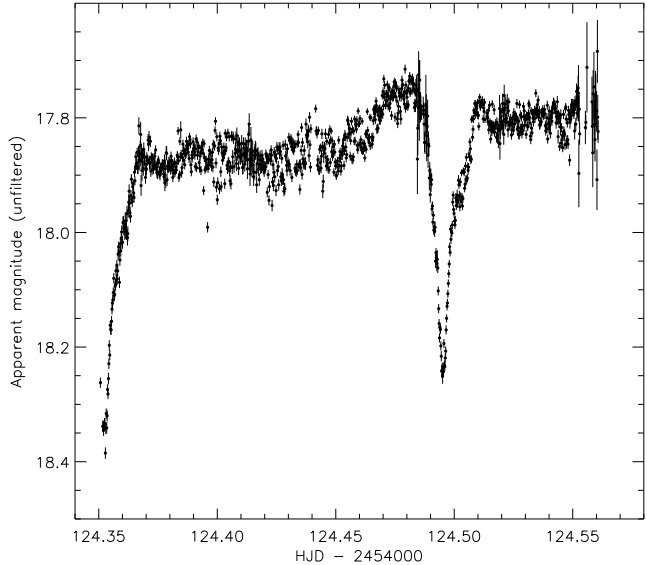}\\
\caption{\label{fig:0754:lcplot} NOT/ALFOSC unfiltered photometry of SDSS\,J0754.}
\end{figure}

We have not used a mass-radius relation for properties of the secondary star for the reasons stated in Paper\,I. In addition, under the current picture of CV evolution, CVs are expected to evolve through an orbital period of 81.5\,min both before and after minimum period. The properties of the secondary star differ significantly between the two different scenarios \citep{Knigge06mn} so cannot be inferred purely from a known orbital period.

The resulting constraints can be seen in Fig.\,\ref{fig:0131:params} and show that the system has a very low mass ratio. If the white dwarf in SDSS\,J0131 has a mass equal to the canonical value of approximately 0.8\Msun\ \citep{SmithDhillon98mn} then the secondary star mass must be below about 0.08\Msun. A more massive white dwarf allows a more massive secondary, but this mass can only be above the substellar limit for an orbital inclination lower than about 35$^\circ$ whereas the double-peaked emission suggests that the inclination is higher than this. SDSS\,J0131 is therefore a good candidate for a CV containing a donor star of a mass appropriate for a brown dwarf.

SDSS\,J0131 strongly resembles GD\,552, which has double-peaked lines and a very similar velocity amplitude of $17 \pm 4$\kms\ \citep{HessmanHopp90aa}. Based on their data and a conviction that the secondary star should not have a substellar mass, \citet{HessmanHopp90aa} suggested that GD\,552 is a low-inclination CV containing a high-mass white dwarf and a main sequence mass donor. However, as short-period CVs are expected to have very low mass secondary components \citep{KolbBaraffe99mn,Politano04apj} then both GD\,552 and SDSS\,J0131 are good candidates for containing substellar mass donors.


\subsection{SDSS J075443.01$+$500729.2}

The discovery spectrum of SDSS\,J0754 was presented by \citet{Szkody+06aj} and shows very weak emission lines at H$\alpha$, H$\beta$, H$\gamma$ and \ion{He}{II} 4686\,\AA\ superimposed on an otherwise featureless blue continuum (Fig.\,\ref{fig:sdssspec}). Szkody et al.\ obtained follow-up polarimetric observations of this object, motivated by the possibility that the 4686\,\AA\ emission indicated a system with a magnetic white dwarf, but found no circular polarisation to a 3\,$\sigma$ upper limit of 0.1\%. The system also was not detected at X-ray wavelengths by the ROSAT satellite.

\begin{table} \begin{center}
\caption{\label{tab:eclipses} Times of eclipses of SDSS\,J0754 and SDSS\,J1555.
Those for SDSS\,J0754 were determined by eye and those for SDSS\,J1555 were
measured by fitting low-order polynomials to the light curve datapoints in the
eclipses. The quoted uncertainties arise mainly from considering the variation
in results from choosing different sets of points to fit for each eclipse. The
calculated values for SDSS\,J1555 give the predicted times from the orbital
ephemeris.}
\begin{tabular}{l c r@{\,$\pm$\,}l r} \hline
Object       & Cycle & \mc{Time of eclipse (HJD)} &   Residual    \\
\hline
SDSS\,J0754  &   -1  &  2454124.3523  &  0.0003   &    0.0008     \\
SDSS\,J0754  &    0  &  2454124.4949  &  0.0003   &    0.0012     \\
SDSS\,J0754  &  118  &  2454141.2302  &  0.0020   &    0.0006     \\
SDSS\,J0754  &  119  &  2454141.3752  &  0.0020   & $-$0.0014     \\
SDSS\,J0754  &  197  &  2454152.5298  &  0.0020   &    0.0005     \\
\hline
SDSS\,J1555  &    0  &  2453873.5943  &  0.0002   &    0.0001     \\
SDSS\,J1555  &    1  &  2453873.6733  &  0.0003   &    0.0001     \\
SDSS\,J1555  &   12  &  2453874.5405  &  0.0002   &    0.0000     \\
\hline \end{tabular} \end{center} \end{table}

Our follow-up photometry of SDSS\,J0754 (Table\,\ref{tab:obslog}, Fig.\,\ref{fig:0754:lcplot}) shows that the system undergoes 0.5\,mag-deep V-shaped eclipses. As the eclipses have sharp minima, we have determined the times of mid-eclipse by eye. For each eclipse, a mirror-image of the light curve was shifted until the two representations of the central parts of the eclipse were in the best possible agreement. The measured times of minimum light and the observed minus calculated values are given in Table\,\ref{tab:eclipses}. The uncertainties in the values are estimated by considering the scatter in the light curve. We have fitted a linear ephemeris to these times of minimum light, finding
\[{\rm Min\,I (HJD}) = 2\,454\,124.3531 (14) + 0.143031 (10) \times E \]
where $E$ is the cycle number and the parenthesed quantities indicate the uncertainty in the last digit of the preceding number. This corresponds to an orbital period of $205.965 \pm 0.014$\,min. The residuals in Table\,\ref{tab:eclipses} are larger than the estimated errors on each measurement. This is not surprising given the properties of this system, and the effect is accounted for in the uncertainties quoted in the above ephemeris.

The shape of the eclipses and the spectral characteristics of the system indicate that it is a high-accretion-rate novalike CV. It was at a similar apparent magnitude during both the NOT and the Tuorla observations. The period from the eclipses is close to the upper edge of the period gap which is observed in the orbital period distribution of the known CVs. The properties of SDSS\,0754 are also consistent with some of the defining characteristics of the SW\,Sextantis class of CV \citep{Thorstensen+91aj,Rodriguez+07mn,Rodriguez+07mn2}.


\subsection{SDSS J155531.99$-$001055.0}

\begin{figure} \includegraphics[width=0.48\textwidth,angle=0]{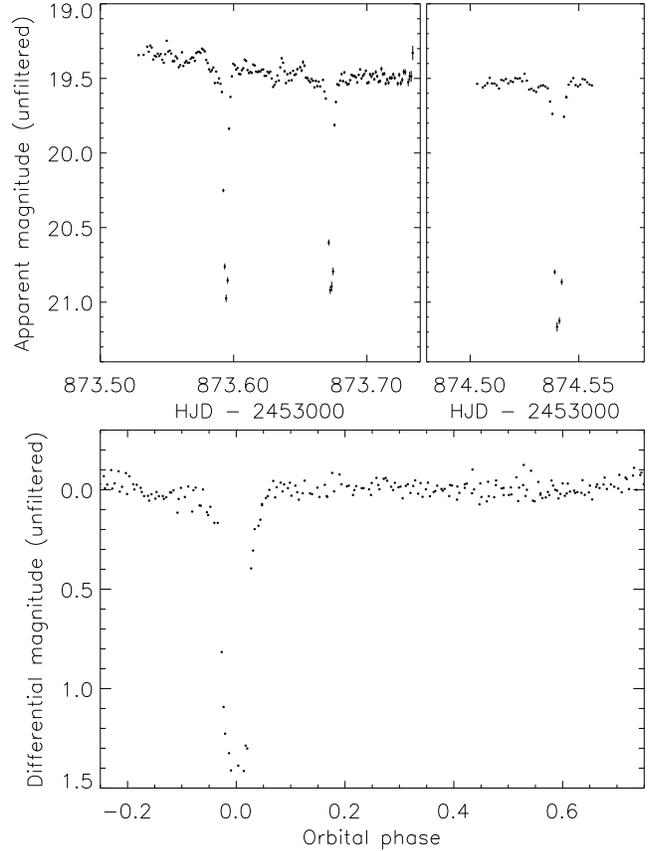}\\
\caption{\label{fig:1555:lcplot} INT/WFC unfiltered photometry of SDSS\,J1555.
Data from the two individual nights are shown in the upper panels. The phased
light curve is shown in the lower panel after subtraction of a low-order
polynomial fit to the outside-eclipse data from each night.} \end{figure}

SDSS\,J1555 was discovered to be a CV by \citet{Szkody+02aj}. Its SDSS spectrum is a little noisy but clearly shows Balmer and \ion{He}{I} emission lines, all of which have the strong double peaks characteristic of accreting binary systems with a high orbital inclination (Fig.\,\ref{fig:sdssspec}). Szkody et al.\ suggested that it may therefore be an eclipsing system. In two nights of INT photometry of this object we detected three eclipses, of depths about 1.6\,mag and with a shape indicating that they are total (Fig.\,\ref{fig:1555:lcplot}).

The brightness variation due to the eclipses is much faster than our photometry, which has a sampling rate of approximately 90\,s. We have therefore measured the eclipse midpoints in two ways: from fitting low-order polynomials to the datapoints in eclipse and by-eye estimates from shifting each eclipse relative to its mirror image. More sophisticated approaches are unwarranted as the light variation is strongly undersampled. The two techniques gave essentially identical results, and we report in Table\,\ref{tab:eclipses} the times of eclipses found by polynomial fittings.

To derive the orbital ephemeris we fitted the polynomial-derived times of eclipse with a linear ephemeris. The observed minus calculated values are given in Table\,\ref{tab:eclipses} and are substantially smaller than the estimated uncertainty in each eclipse timing. Considering the available data, and the simplicity of our methods, we prefer to attribute this agreement to good fortune rather than an indication that our errors are overestimated. The resulting ephemeris (with our conservative measurement errors retained) is
\[{\rm Min\,I (HJD}) = 2\,453\,873.59434(17) + 0.078847(22) \times E \]

\begin{figure} \includegraphics[width=0.48\textwidth,angle=0]{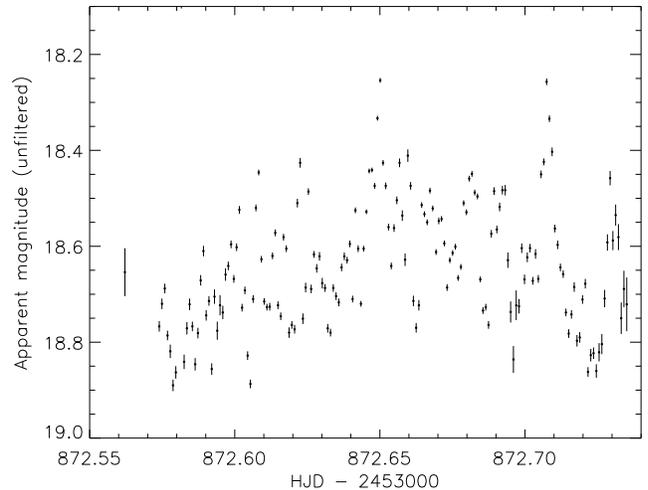}\\
\caption{\label{fig:1656:lcplot} INT unfiltered photometry of SDSS\,J1656.} \end{figure}

SDSS\,J1555 is therefore another object identified as a CV by the SDSS and with a period below the gap, in this case $113.54 \pm 0.03$\,min. Because it undergoes deep eclipses it is also a very promising candidate for the accurate measurement of the masses and radii of its component stars \citep{Wood+86mn,Littlefair+06mn,Littlefair+06sci}.


\subsection{SDSS J165658.12+212139.3}                                                         \label{sec:1656}

This was found to be a CV by \citet{Szkody+05aj} from the presence of double-peaked Balmer and \ion{He}{I} line emission (Fig.\,\ref{fig:whtspec}). It has been detected as an X-ray source by ROSAT \citep{Voges+99aa,Voges+00iauc}. Our four hours of INT/WFC photometry (Fig.\,\ref{fig:1656:lcplot}) indicate that this system is variable but with no coherent period. This flickering behaviour is present in many CVs and is generally observed to arise from either the hot spot on the outer edge of the disc, or the boundary layer where the inner disc encounters the surface of the white dwarf \citep{Bruch92aa,Bruch00aa}. This suggests that the period of SDSS\,J1656 will need to be found spectroscopically rather than photometrically.


\subsection{SDSS J205914.87$-$061220.4}                                                       \label{sec:2059}

\begin{figure} \includegraphics[width=0.48\textwidth,angle=0]{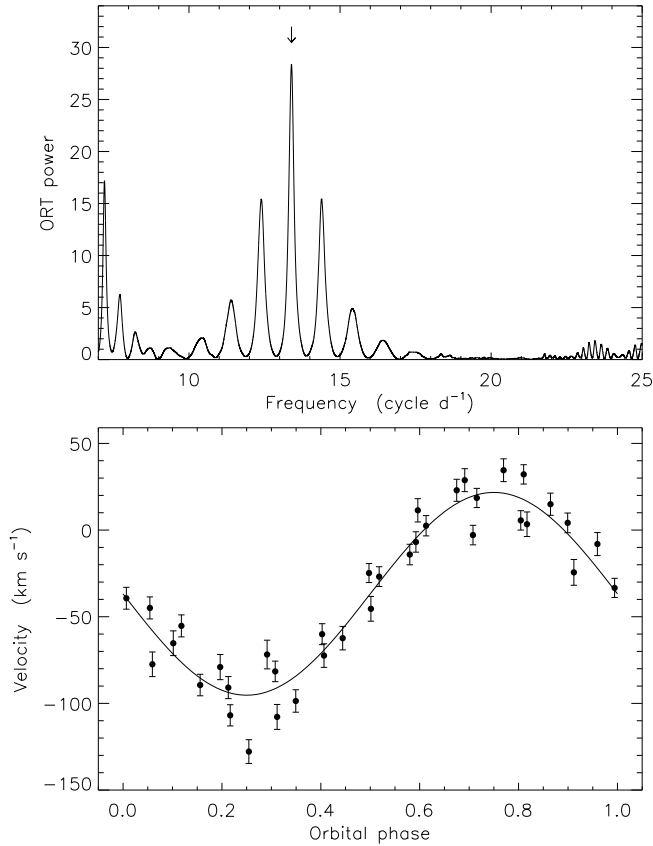}\\
\caption{\label{fig:2059:rvplot} ORT periodogram (top) and phased radial velocity
curve (bottom) for SDSS\,J2059. The radial velocities were measured from the
H$\alpha$ emission lines in our WHT/ISIS spectra using the double Gaussian
technique (see text for details). The frequency corresponding to the best-fitting
orbital period from {\sc sbop} is indicated with a downward-pointing arrow.} \end{figure}

\begin{figure} \includegraphics[width=0.48\textwidth,angle=0]{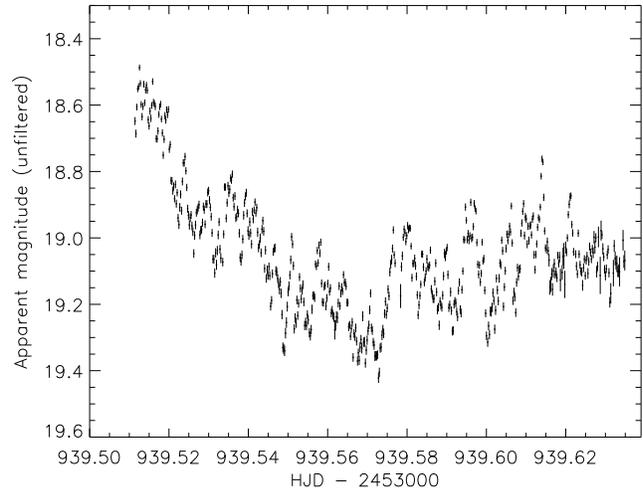}\\
\caption{\label{fig:2059:lcplot} NOT unfiltered photometry of SDSS\,J2059.} \end{figure}

The SDSS discovery spectrum of SDSS\,J2059 was presented by \citet{Szkody+03aj} and, like SDSS\,J0131, it is typical of quiescent dwarf novae. The Balmer emission lines are strong and emission occurs for the Paschen lines, \ion{He}{I}, \ion{Ca}{II} and \ion{Fe}{II}. Weak emission is also present at \ion{He}{II} 4686\,\AA. Unlike SDSS\,J0131 there is no clear sign of wide Balmer absorption wings arising from the white dwarf.

We obtained a total of 38 WHT/ISIS spectra of SDSS\,J2059 over two nights. This CV was in outburst during the first night, with a visual magnitude of approximately 16.6 shortly after our observations (D.\ Boyd, private communication). During the second night it faded from magnitude 17.0 to 17.5 (unfiltered; T.\ Krajci, private communication), which compares to an apparent magnitude of $g = 18.38$ in the SDSS imaging observations. To our knowledge this is the first outburst recorded for SDSS\,J2059, but the available observations only cover the post-maximum decline so we have no knowledge of the duration or peak magnitude of the outburst.

In Fig.\,\ref{fig:whtspec} we show the average spectrum from each night. The flux in the emission lines is similar on the two nights but the average spectrum from the first night shows strong Balmer and \ion{He}{I} 4471\,\AA\ absorption from an optically thick disc. This is narrower than the Balmer absorption normally seen from white dwarfs (which can be noted by comparing it to the spectrum of SDSS\,J0131 on the same plot) and is characteristic of disc absorption with a smaller pressure broadening than for white dwarfs. The average spectra of SDSS\,J2059 on the two nights are typical for CVs which are declining from outburst \citep[e.g.,][]{Hessman+84apj,MoralesMarsh02mn}.

We measured radial velocities from our WHT spectra in the same way as for SDSS\,J0131, and similarly to this CV found that the best results were obtained for the H$\alpha$ emission line using double Gaussians of FWHM 300\kms\ and separation 1000\kms. The AoV and ORT periodograms both have a strong signal at 107.6\,min (Fig.\,\ref{fig:2059:rvplot}), with the ORT periodogram in particular having only weak one-day aliases (see Table\,\ref{tab:orbits}). This signal remains strong for a wide range of double-Gaussian separations (600 to 2000\kms) and for the single Gaussian technique. Bootstrapping simulations give a probability of at least 98\% that it corresponds to the actual period (Table\,\ref{tab:boot}). Fitting a sine curve to the best radial velocities (H$\alpha$, double Gaussian with FWHMs 300\kms\ and separation 1000\kms) gives a period of $107.52 \pm 0.14$\,min (Table\,\ref{fig:2059:rvplot}). Trailed phase-binned spectra are shown in Fig.\,\ref{fig:trailed} for each night separately.

As SDSS\,J2059 had a much brighter accretion disc on the first night than the second night its spectral characteristics are quite different. We have been careful to ensure that this has not affected our radial velocity measurements. Radial velocities determined using cross-correlation with a single Gaussian in fact give a very similar periodogram to our double-Gaussian values. As the single Gaussian technique is essentially unaffected by the changing shape of the spectrum outside the emission lines, this indicates that the double-Gaussian method has remained reliable in this case.

Three hours of unfiltered photometry of SDSS\,J2059 were obtained with the NOT several months before our observations (Fig.\,\ref{fig:2059:lcplot}). These show the flickering characteristic of mass transfer in binary stars, but do not show any variation at the orbital period.


\subsection{SDSS J210449.94+010545.8}                                                           \label{sec:2104}

\begin{figure} \includegraphics[width=0.48\textwidth,angle=0]{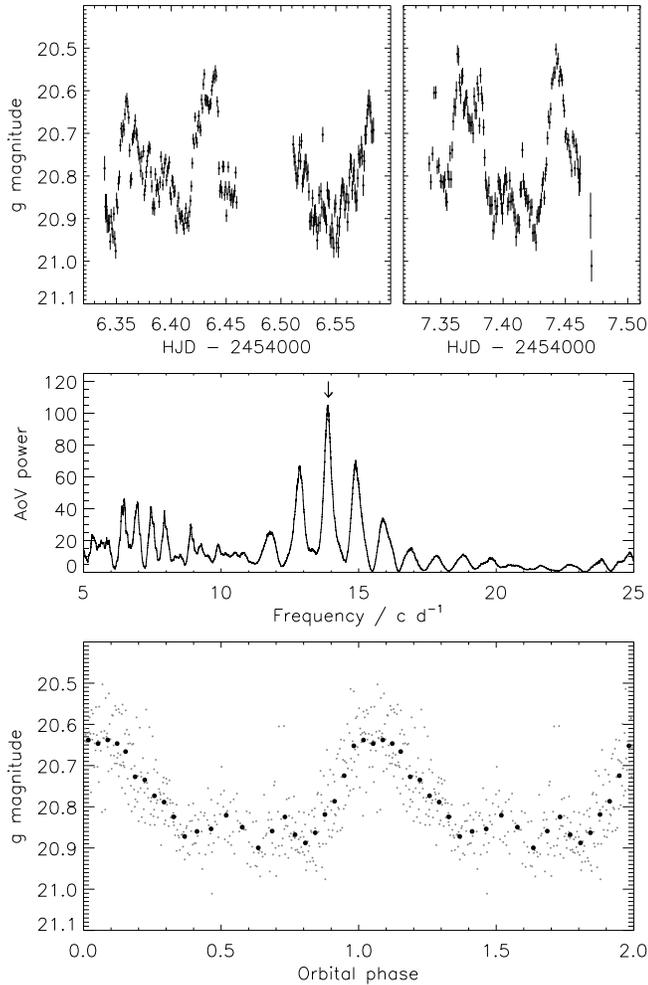}\\
\caption{\label{fig:2104:lcplot} WHT imaging photometry of SDSS\,J2104. The
data were obtained using an SDSS $g$ filter. The upper panels show the observed
light curves and the middle panel shows the AoV periodogram of the data. The lower
panel shows the phased data (grey points) and phasebinned data (black points).}
\end{figure}

SDSS\,J2104 was discovered to be a CV by \citet{Szkody+06aj}. Its spectrum shows weak emission lines at H$\alpha$ and H$\beta$ and vague hints of the \ion{He}{I} emission found in almost all CVs. The H$\alpha$ emission line seems to be triple-peaked, but this would need confirmation from a spectrum with a higher ratio of signal to noise. Szkody et al.\ noted that the system was substantially brighter at the epoch of the SDSS imaging observations than when its SDSS spectrum was obtained, and that it is faint on Digital Sky Survey plates. They concluded that it is a dwarf nova which was not in quiescence when the SDSS $ugriz$ magnitudes were obtained. \reff{At this point it was at $g = 17.23$, whereas by convolving its flux-calibrated spectrum with a $g$ response function we find a magnitude of $g = 20.11$ at its spectroscopic epoch.}

We expected SDSS\,J2104 to be a short-period system which therefore exhibits large outbursts for only a small fraction of the time, so would normally be too faint to be studied using the WHT/ISIS. We therefore used the WHT Auxiliary Port to obtain approximately 350 imaging observations of it through the SDSS $g$ filter. The small field of view of the WHT Aux Port CCD unfortunately meant that only a single comparison star was available, which was substantially fainter than SDSS\,J2104, lowering the quality of our photometry.

The light curve (Fig.\,\ref{fig:2104:lcplot}) has a mean magnitude of $g = 20.8$, making SDSS\,J2104 3.6\,mag fainter than at the time of the SDSS imaging. It also demonstrates a clear orbital hump which shows up nicely in the AoV and ORT periodograms (see Fig.\,\ref{fig:2104:lcplot} for the ORT periodogram). Fitting a sine curve to the data -- an acceptable approximation to the light variation -- gives a period of $103.62 \pm 0.12$\,min. The one-day aliases on each side have a much weaker power than the main peak and so can safely be discounted -- our bootstrapping simulations give a probability of 99.8\% that the main peak is the correct alias (Table\,\ref{tab:boot}).


\section{Discussion and conclusions}                             \label{sec:conclusion} \label{sec:discussion}

The majority of CVs have been discovered through optical variability or X-ray emission, meaning that they constitute a very biased sample of objects \citep{Gansicke05aspc}. The CVs discovered by the SDSS have been identified spectroscopically (after targeting due to their $ugriz$ colour indices), and the faint limiting magnitude of this survey means that this sample is much less biased against the intrinsically faint and short-period objects which are predicted to comprise the vast majority of the intrinsic population of CVs. Only one of the CVs studied in this work (SDSS\,J1656) is known to be an X-ray source.

As the SDSS observes away from the Galactic plane and has a fairly faint limiting magnitude, it gives an approximately volume-limited sample for objects brighter than a certain apparent magnitude. To calculate this magnitude we have assumed that a distance of three scale heights above the Galactic plane (which includes 95\% of thin-disc objects) is the lower limit for delivering a volume-limited sample. Adopting a scale height of 190\,pc \citep[e.g.][]{Patterson84apjs} and a minimum Galactic latitude for the main SDSS survey area of $|b| > 30^\circ$ \citep{York+00aj} gives a minimum distance of 1140\,pc and so a minimum distance modulus of 10.3. As the SDSS CV sample is approximately complete to $g = 20.5$, it is volume-limited for objects with absolute magnitude $M_g \ga 10.2$.

Whilst there is a clear relation between absolute magnitude and orbital period for dwarf novae \citep[see][]{Thorstensen03aj}, we cannot use this as it does not extend to all classes of CV. Under the assumption that U\,Gem ($M_V \approx 10$), a dwarf nova with a low mass transfer rate, represents the faint limit for CVs with $P \ga 3$\,hr \citep{Harrison+99apj}, we find that the SDSS sample of CVs is therefore complete for all such systems with periods longer than the period gap. It is substantially incomplete for CVs with periods shorter than the period gap, e.g., SDSS\,J0131 has an absolute magnitude of $M_g = 10.7 \pm 0.6$. Also, magnetic CVs can be fainter as their accretion discs are truncated or entirely missing. An example is the prototype system, AM\,Her, which has $M_V \approx 11$ \cite{Thorstensen03aj}, and a counter-example is the short-period system EX\,Hya, which has $M_V \approx 9.0$ in quiescence \citep{Beuermann+03aa}. Conversely, novalikes have very luminous accretion discs so have bright absolute magnitudes; those examples closer to the Earth will also be missed by the SDSS due to its bright limit of $g = 15.0$ for spectroscopic observations \citep{Stoughton+02aj}. A full investigation of the observational selection effects present in the CV sample is in progress.

\begin{table} \begin{center}
\caption{\label{tab:result} Summary of the orbital
periods obtained for the objects studied in this work.}
\begin{tabular}{l r@{.}l@{\,$\pm$\,}r@{.}l l} \hline
Object      & \multicolumn{4}{c}{Period (min)} & Notes \\ \hline
SDSS\,J0050 &   80&3   & 2&2    & WHT photometry      \\
SDSS\,J0131 &   81&54  & 0&13   & WHT spectroscopy    \\
SDSS\,J0754 &  205&965 & 0&014  & Eclipsing           \\
SDSS\,J1555 &  113&54  & 0&03   & Eclipsing           \\
SDSS\,J1656 &  \mc{ }  & \mc{ } & {\it flickering}    \\
SDSS\,J2059 &  107&52  & 0&14   & WHT spectroscopy    \\
SDSS\,J2104 &  103&62  & 0&12   & WHT photometry      \\
\hline \end{tabular} \end{center} \end{table}

We are engaged in a project to categorise the sample of CVs discovered by the SDSS, the primary goal being measurement of the orbital period and other available quantities for each object. In this work we have presented spectroscopy and photometry from four telescopes, and measured the orbital periods of six SDSS CVs (Table\,\ref{tab:result}).

Of the six CVs for which we have measured orbital periods, two have values close to the minimum period for CVs with unevolved secondary stars: SDSS\,J0050 with 80.3\,min and SDSS\,J0131 with 80.54\,min. Three objects have slightly longer periods but still qualify as short-period systems: SDSS\,J1555 with 113.54\,min, SDSS\,J2059 with 107.518\,min and SDSS\,J2104 with 103.62\,min. We observed SDSS\,J2059 during the decline from its first recorded outburst so are able to classify it as a dwarf nova. Only one system, SDSS\,J0754, has an orbital period ($205.965 \pm 0.014$\,min) which is longer than the 2--3\,hr period gap. It also has a high mass transfer rate.

The orbital velocity amplitude for SDSS\,J0131 is only $19.6 \pm 1.4$\kms. Using several constraints we find that this system has an extremely low mass ratio and contains a massive white dwarf and/or a very low-mass secondary star. The mass donor is a good candidate for being a CV secondary star with a mass below the minimum required for core hydrogen burning.

Our photometry of SDSS\,J0754 and SDSS\,J1555 show that they are eclipsing systems with 0.5\,mag and 1.5\,mag deep eclipses (in unfiltered light), respectively. SDSS\,J1555 is a good candidate for accurate measurement of the absolute properties of a short-period CV, as the functional form of its eclipse is well understood \citep[e.g.][]{Horne++91apj,Steeghs+03mn,Littlefair+06sci}. A similar analysis for SDSS\,J0754 is not currently possible due to the complexities of the novalike class of CVs.

A recurrent theme in the previous paragraphs is that the orbital periods we are finding for the SDSS CVs are almost all shorter than the 2--3\,hr period gap, and in many cases adjacent to the observed 76\,min minimum period. This is consistent with the expectation that shorter-period CVs are not only intrinsically fainter but more populous that their longer-period counterparts \citep{Patterson++05pasp}. Moreover, the orbital period distribution of those SDSS CVs with known periods is much closer to theoretical predictions than the distribution for the complete sample of known CVs.


\section*{Acknowledgements}

We would like to warmly thank David Boyd and Tom Krajci for their speed in responding to our request for observations of SDSS\,J2059 after our spectroscopic observations indicated that it might be in outburst. We are  also grateful for a positive and timely review from the anonymous referee.

The reduced spectra and radial velocity observations presented in this work will be made available at the CDS ({\tt http://cdsweb.u-strasbg.fr/}) and at {\tt http://www.astro.keele.ac.uk/$\sim$jkt/}.

This work is based on observations made with the William Herschel Telescope and the Isaac Newton Telescope, which are operated on the island of La Palma by the Isaac Newton Group in the Spanish Observatorio del Roque de los Muchachos (ORM) of the Instituto de Astrof\'{\i}sica de Canarias (IAC).
Observations were also made with the Nordic Optical Telescope, operated on the island of La Palma jointly by Denmark, Finland, Iceland, Norway, and Sweden, also in the ORM of the IAC. These observations used ALFOSC, which is owned by the Instituto de Astrof\'{\i}sica de Andaluc\'{\i}a (IAA) and operated at the Nordic Optical Telescope under agreement between IAA and the NBIfAFG of the Astronomical Observatory of Copenhagen.

JS acknowledges financial support from STFC in the form of a postdoctoral research assistant position. BTG acknowledges financial support from PPARC in the form of an advanced fellowship. TRM was supported by a PPARC senior fellowship during the course of this work.
DDM DdM acknowledgs financial support by ASI and INAF via grant I/023/05/0.
AA thanks the Royal Thai Government for a studentship.

The following internet-based resources were used in research for this paper: the ESO Digitized Sky Survey; the NASA Astrophysics Data System; the SIMBAD database operated at CDS, Strasbourg, France; and the ar$\chi$iv scientific paper preprint service operated by Cornell University.

Funding for the Sloan Digital Sky Survey (SDSS) has been provided by the Alfred P.\ Sloan Foundation, the Participating Institutions, the National Aeronautics and Space Administration, the National Science Foundation, the U.\ S.\ Department of Energy, the Japanese Monbukagakusho, and the Max Planck Society. The SDSS website is {\tt http://www.sdss.org/}.


\bibliographystyle{mn_new}


\bsp

\label{lastpage}

\end{document}